# ImageBox3: No-Server Tile Serving to Traverse Whole Slide Images on the Web

Praphulla MS Bhawsar[1], Erich Bremer[2], Máire A Duggan[3], Stephen Chanock[1], Montserrat Garcia-Closas[1], Joel Saltz[2], Jonas S Almeida[1]

[1] - Division of Cancer Epidemiology and Genetics, National Cancer Institute, National Institutes of Health, Maryland, USA
[2] - Department of Biomedical Informatics, Stony Brook University, Stony Brook, NY, USA
[3] – Department of Pathology and Laboratory Medicine, Cumming School of Medicine, University of Calgary

## Abstract
Whole slide imaging (WSI) has become the primary modality for digital pathology data. However, due to the size and high-resolution nature of these images, they are generally only accessed in smaller sections or "tiles" via specialized platforms, most of which require extensive setup and/or costly infrastructure. These platforms typically also need a copy of the images to be locally available to them, potentially causing issues with data governance and provenance. To address these concerns, we developed ImageBox3, an in-browser tiling mechanism to enable zero-footprint traversal of remote WSI data. All computation is performed client-side without compromising user governance, operating public and private images alike as long as the storage service supports HTTP range requests (standard in Cloud storage and most web servers). ImageBox3 thus removes significant hurdles to WSI operation and effective collaboration, allowing for the sort of democratized analytical tools needed to establish participative, FAIR digital pathology data commons.

Availability:
code - https://github.com/episphere/imagebox3 ;
fig1 (live) - https://episphere.github.io/imagebox3/demo/scriptTag ;
fig2 (live) - https://episphere.github.io/imagebox3/demo/serviceWorker ;
fig 3 (live) - https://observablehq.com/@prafulb/imagebox3-in-observable .

## 1. Introduction
For 6 years[1,2] we have been advancing computational solutions that progressively facilitate the analysis of large digital pathology images by operating them remotely in place (i.e., operating on the part without downloading the whole). This is in contrast with existing platforms like QuPath[3], caMicroscope[4] and others[5,6] that rely on server-side components to perform any data parceling or computation, using the client as merely a place to render the results of the computation. While this approach is understandable historically given the scale of WSI data and the resource limitations of the client machine, it comes with serious issues. Just to view a whole slide image for instance, it is necessary to use specialized software applications, often needing expensive hardware and incurring considerable licensing fees for use on an institutional scale. Moreover, to maintain governance, the images have to be made locally available to the platform, leading to data duplication and provenance issues. Interoperability is a hard problem on these platforms, making FAIR-ness[7] becomes hard to realize. External collaboration is similarly tedious to achieve given the siloed nature of these software. In essence, these platforms may be described as enclaves that provide useful functionalities but at the cost of the data (and the analyses) being locked inside.

As part of our efforts to tackle these problems, we first showed how the use of the cloud for file storage could offer a robust, managed solution for ensuring data governance and regulatory compliance[1]. Next, we demonstrated ImageBox2, a containerized tile server to mediate calls to remote whole slide images using HTTP range requests[2]. While ImageBox2 did provide a means to retrieve sections or tiles from remote WSI data, it still relied on a proxy server that would need to be deployed and maintained. The current work represents the culmination of our efforts, in that we eliminate the intermediary altogether. We describe ImageBox3, an implementation where the tiling operation is absorbed as an entirely in-browser process. Since no proxies or server-side components are needed anymore, the problem of accessing WSI data, both local and remote, is reduced to a zero-footprint solution. The image analysis application can thus engage the target image entirely within the governance of the user on the user's own machine. Notably, this remote operation enables the dissemination of analytics procedures, from simple image processing to deep learned Artificial Intelligence (AI) models, without ever having to move or copy the data, while simultaneously allowing for dramatic improvements in scalability and portability.

The extended period of time needed to achieve this result reflects as much the intricacies of the tiling structures of digital pathology images, as it does the development of Web Computing itself. Three factors in particular created the opportunity for a zero-footprint solution for WSI tiling. First and foremost, the growing use of Cloud Computing infrastructure extensively commoditized the availability of modern web servers to the point where most WSI data, public or private, can now be accessed securely and on demand via HTTP range requests[8]. Secondly, Web standards have furthered support for functionalities that were previously the exclusive province of native software applications. Amongst those, support for multi-threaded in-browser computation[9] is among the most transformative. Thirdly, as digital pathology and other disciplines using large images tend to use similar file formats, it is possible to apply the work done for images of one kind to those of another. In particular, the availability of GeoTIFF.js, a library for operating on cloud-optimized GeoTIFF files in the browser[10], allowed us to reuse the code produced by the geospatial community for operating on WSI data.

In summary, our approach advances a reading of the evolution of modern computing as corresponding to the "client-ification" of the server. The reverse (the "server-ification" of the client) would be a harder argument to defend since a critical advantage of web computing is the security and governance of the sandbox where it operates. As a data scientist might put it, computing in the web browser is far FAIR-er. Hence the title of this report and the simplest description of the open source ImageBox3 library: a zero-footprint, "no-server tile server" for operating large whole slide images.

## 2. Results

As described before, ImageBox3 leverages the similarities in image formats used for geospatial data and WSI. Although WSI formats tend to be proprietary to the scanner vendors, the most common ones are based on the Tag Image File Format (TIFF)[11]. Cloud-optimized GeoTIFFs[12] use the same backbone; as a result, any program that can traverse GeoTIFF data can be extended for WSI data without much effort. ImageBox3 builds on the foundation laid by the GeoTIFF.js library to deliver on-the-fly tiling of remote whole slide images. To support interoperability across platforms, ImageBox3 follows the IIIF3.0 standard[13] to parameterize requests for tiles.

The ability to retrieve individual tiles from remote images thus ultimately depends on support for HTTP range requests (retrieving a subset of the remote volume) by the file server. As noted above, the wide availability of modern web servers behind public facing content and services is a development largely associated with the nearly universal reliance on Cloud Computing infrastructure, even to host legacy systems. Accordingly, the "no-server tile server" functionality explored advances the use of range requests with three distinct software constructs. Each of them is accompanied by an openly accessible in-browser implementation, entirely in JavaScript.

### 2.1 Direct tilling

A straightforward implementation that makes direct range-request HTTP calls from the main thread to a remote volume. This process can be illustrated and verified by pointing this application to a remote WSI image so as to retrieve tiles at various locations and magnifications. (Fig1)

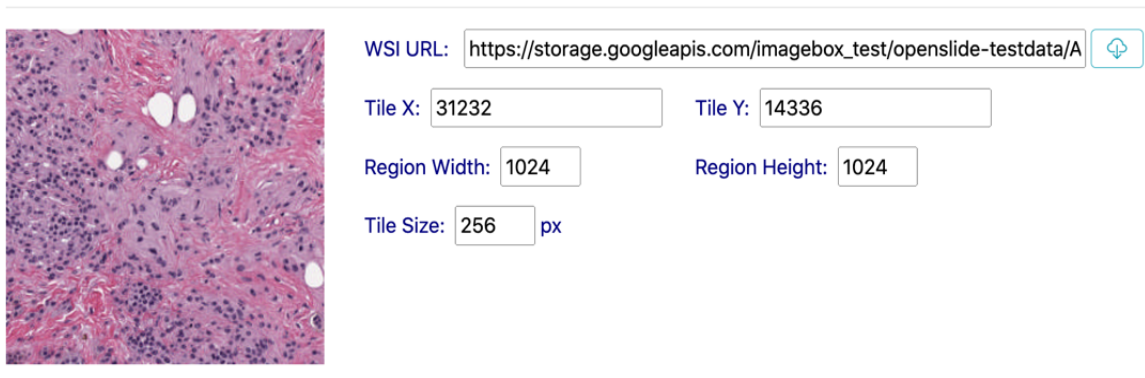

Fig1 – Direct titling demonstration at https://episphere.github.io/imagebox3/demo/scriptTag. Note links to source code and issues submission.
A non-server implementation, which makes direct HTTP byte range requests and performs all decoding and rendering computation on the main thread.

## 2.2 Embedded tilling

An embedded WSI viewer that mediates calls to tiles by extending existing high-resolution image viewers such as OpenSeadragon[14]. This solution relies on service workers[15], a web serving functionality supported by all modern web browsers. The implementation is illustrated and verified by this implementation (Fig 2), where for all practical purposes the embedded in-browser server replaces the conventional server-side equivalent.

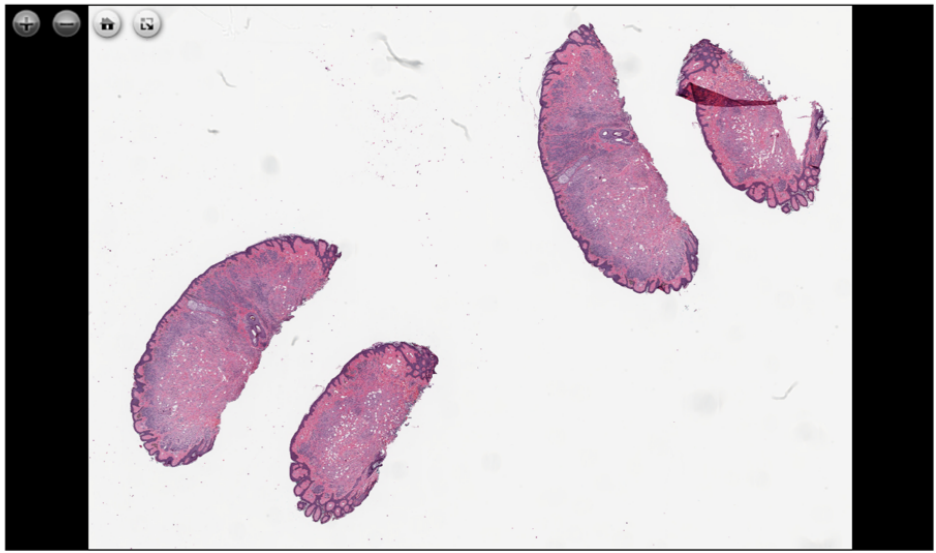

Fig2 - Demonstration at https://episphere.github.io/imagebox3/demo/serviceWorker using the OpenSeadragon viewer. This implementation relies on a service worker, a native pseudo-serving functionality of the web browser, to handle requests for tiles made by the viewer. These parameterized requests are translated to the appropriate byte range by ImageBox3, processed as individual tiles and served back to the viewer.

## 2.3 Software development kit (SDK)

A supporting JavaScript SDK underlying both #1 and #2. In addition to enabling software development in general, the SDK is also particularly suitable for notebook uses where it is simply imported, like any other library. The latter is illustrated here without breaking the zero-footprint promise, with a reactive, reproducible and collaborative Observable notebook (Fig 3).

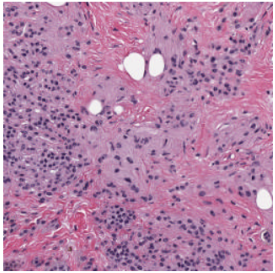

Fig3 – Demonstration at https://observablehq.com/@prafulb/imagebox3-in-observable of importing and using the ImageBox3 SDK into a reactive notebook. This example is only fully understood by engaging and modifying the programmatic components. To illustrate and invite that engagement, the code importing imageBox3 SDK in this snapshot is exposed by clicking the arrow tip.

## 3. Discussion

The ability to retrieve tiles of any size and magnification from large whole slide images, with no additional effort than to deposit the images in cloud storage, has the potential to reframe the stewardship of digital pathology data towards FAIR data commons. We argue here for two specific features of more data-centric image repositories, those of disintermediated operation and robust governance, by first revisiting what disintermediated tile serving will do for each of the 4 letters of FAIR[7, 16].

**Findable** - At its most basic, direct calling renders the dereferenceable URL of the full image volume, also its fully self-described Universal Resource Identifier (URI). In contrast, when mediation by a repository-specific API is involved in the dereferencing, the identifier will likely be that of an index to the image volume.

**Accessible** - Cloud storage services invariably provide the image submitter with a range of options to manage access. These often include the full range of mechanisms, from token-based simple auth to advanced OAuth2.0-mediated management of users, groups and roles. This is in stark contrast to current repositories where access management has to be integrated with the operation of local tilling processes. Instead, with ImageBox3 direct tiling, the API call to the full image URL simply includes the token recognized by an agreeable cloud-hosted identity service such as Google Identity service or Login.gov as authorized by the user.

**Interoperable** - A unified interoperability model may well be the most impactful advantage of ImageBox3's disintermediated tiling retrieval: the tiling scheme is specific to each image format. Consequently, the parameterization of the URL call by a range request header only requires the client application to know what format the image was encoded in in the first place. This completely removes the additional complication of learning how a specific repository parameterizes the operation of server-side computed tile rendering. And as an added bonus, Web Computing itself is now also available beyond the browser in environments traditionally preferred by data scientists, such as via the V8 package in R[17] or Pyodide for Python[18]. This means that the ImageBox3 tiling mechanism can also be engaged in those environments without much hassle.

**Reusable** - A growing number of image repositories are constantly being developed. Each will often advance a competing recommendation for a universal interoperability model. Similarly, a growing number of analytical applications seek to support federated learning approaches where instead of aggregating the data, the model parameters (typically through deep learning) are identified by the distributed iteration of the regression process[19]. This is not a portable or scalable arrangement, where both the interoperability model and the data traversal procedure have to be configured for each combination of the two to work. Even without considering additional barriers to reproducibility, such as the uncertain durability of a repository-specific solution, the image submitter/custodian will typically not be involved in governing it, hindering trust in the process. As discussed for Accessibility and Reusability above, by orchestrating tiling directly from the analytical environment, either Cloud-hosted or on edge devices (including web browsers), we make the entire process both predictable and transparent to the data provider, as well as to the consumer of the analytic results. This transparency in turn facilitates easy reusability and extensibility of those results and the computational artifacts that were used to obtain them.

The ImageBox3 implementation at https://github.com/episphere/imagebox3 is at its core a proof-of-concept that validates browser-based operation of remote WSI data. There are several functionalities that would need to be added to the library to support the operations typically used in a WSI analysis pipeline. For instance, the library has only been tested to support Generic TIFFs and Aperio SVS images, and only those encoded with JPEG compression. With the development of WebAssembly, however, it might not be unreasonable to expect a compiled interface to various WSI formats in the near future, similar to or even derived from the OpenSlide[20] or BioFormats[21] libraries. Indeed, a new generation of file formats amenable to remote operation is emerging to increase FAIR-ness of bioimaging [22].

In conclusion, a software engineering solution was identified, implemented and made publicly available in the public domain to traverse large Whole Slide Images (WSI) directly from where they are stored. No proxy servers or services are needed to mediate the process. In addition to the logistic simplification of the process of analyzing images, this enables a far safer approach to distributing computationally intensive analytics such as AI over the web[23]. Furthermore, ImageBox3 enables a fundamentally new approach to the development of Image Commons resources. The latter is of particular relevance for the FAIR stewardship of scientific data[7, 16]. In many regards the direct tiling route taken by ImageBox3 is as good of a match to the original idea of data commons as an API ecosystem[24]. At its most basic, the only API needed is the HTTP call to the Universal Resource Locator (URL) to the raw file with a range request parameter.